\documentclass{article}
\usepackage[preprint]{spconf}
\usepackage{amsmath,graphicx}
\usepackage{amssymb}
\usepackage{subcaption}
\usepackage[hidelinks]{hyperref}
\usepackage{url}
\usepackage{todonotes}
\usepackage{siunitx}
\usepackage[T1]{fontenc}
\usepackage[utf8]{inputenc}

\usepackage{tikz}
\usetikzlibrary{shapes, arrows, positioning, decorations.pathreplacing, calc, shadows}

\DeclareMathOperator{\Var}{Var}
\let\originalleft\left
\let\originalright\right
\renewcommand{\left}{\mathopen{}\mathclose\bgroup\originalleft}
\renewcommand{\right}{\aftergroup\egroup\originalright}

\DeclareMathOperator*{\argmin}{arg\,min}
\makeatletter
\newcommand{\definetrim}[2]{%
  \define@key{Gin}{#1}[]{\setkeys{Gin}{trim=#2,clip}}%
}
\makeatother
\usepackage{bm} %
\usepackage{braket} %
\newcommand{\vect}[1]{\bm{\mathrm{#1}}}
\newcommand{\norm}[1]{\lVert#1\rVert_{2}}
\newcommand{\card}[1]{\vert#1\vert}
\newcommand{\avg}[1]{\braket{#1}}
\newcommand{\ngb}{\mathcal{N}}
\newcommand{\N}[2]{\ngb_{#1}(#2)}
\newcommand{\X}{X}
\newcommand{\x}{\vect{x}}
\newcommand{\n}{n}
\newcommand{\Xtilde}{\tilde{X}}
\newcommand{\xtilde}{\vect{\tilde{x}}}
\newcommand{\G}{G}
\newcommand{\g}{\vect{g}}
\newcommand{\nprime}{n^\prime}
\newcommand{\w}{w}
\newcommand{\h}{h}
\newcommand{\f}{f}
\newcommand{\fe}{f_{e}}
\newcommand{\fd}{f_{d}}
\newcommand{\y}{\vect{y}}
\newcommand{\Loss}{\mathcal{L}}
\DeclareMathOperator{\FL}{FL}
\DeclareMathOperator{\dch}{d_{ch}}
\DeclareMathOperator{\drep}{d_{rep}}
\newcommand{\idw}{\omega}
\newcommand{\Idw}{\Omega}
\newcommand{\idwn}{\hat{\omega}}
\newcommand{\pgrid}[1][]{\vect{p}_{\mathrm{grid}#1}}
\newcommand{\ppush}[1][]{\vect{p}_{\mathrm{push}#1}}
\newcommand{\ppull}[1][]{\vect{p}_{\mathrm{pull}#1}}
\newcommand{\mXXt}{m_{\X \to \Xtilde}}
\newcommand{\mXtX}{m_{\Xtilde \to \X}}

\newcommand{\occ}{o}

\title{Folding-based compression of point cloud attributes}
\name{Maurice Quach \qquad Giuseppe Valenzise \qquad Frederic Dufaux\thanks{Funded by the ANR ReVeRy national fund (REVERY ANR-17-CE23-0020).}}
\address{Universit\'e Paris-Saclay, CNRS, CentraleSup\'elec, Laboratoire des signaux et syst\`emes\\
91190 Gif-sur-Yvette, France}
\copyrightnotice{Copyright 2020 IEEE}
\toappear{ICIP 2020}

\begin{document}
\maketitle
\begin{abstract}
Existing techniques to compress point cloud attributes leverage either geometric or video-based compression tools.
We explore a radically different approach inspired by recent advances in point cloud representation learning.
Point clouds can be interpreted as 2D manifolds in 3D space.
Specifically, we fold a 2D grid onto a point cloud and we map attributes from the point cloud onto the folded 2D grid using a novel optimized mapping method.
This mapping results in an image, which opens a way to apply existing image processing techniques on point cloud attributes.
However, as this mapping process is lossy in nature, we propose several strategies to refine it so that attributes can be mapped to the 2D grid with minimal distortion.
Moreover, this approach can be flexibly applied to point cloud patches in order to better adapt to local geometric complexity.
In this work, we consider point cloud attribute compression; thus, we compress this image with a conventional 2D image codec.
Our preliminary results show that the proposed folding-based coding scheme can already reach performance similar to the latest MPEG Geometry-based PCC (G-PCC) codec.
\end{abstract}


\section{Introduction}
\label{sec:intro}

A point cloud is a set of points in 3D space which can have associated attributes such as color or normals.
Point clouds are essential for numerous applications ranging from archeology and architecture to virtual and mixed reality.
Since they can contain millions of points with complex attributes,  efficient point cloud compression (PCC) is essential to make these applications feasible in practice.

When compressing a point cloud, we usually consider two aspects: the geometry, that is the 3D coordinates of each individual point, and the attributes, for example RGB colors.
Moreover, we can differentiate dynamic point clouds, which change in the temporal dimension, from static point clouds.
The Moving Picture Experts Group (MPEG) is leading PCC standardization efforts \cite{schwarz_emerging_2018}.
Specifically, two main solutions have emerged. The first one, Geometry-based PCC (G-PCC), uses native 3D data structures, while the second one, Video-based PCC (V-PCC), targets mainly dynamic point clouds, and projects the data on a 2D plane to make use of available video codecs such as HEVC.

Point clouds can be interpreted as 2D discrete manifolds in 3D space.
Therefore, instead of compressing point cloud attributes using 3D structures such as octrees, we can fold this 2D manifold onto an image.
This opens many avenues of research, as it provides, e.g., a way to apply existing image processing techniques straightforwardly on point cloud attributes.
In this work, we propose a novel system for folding a point cloud and mapping its attributes to a 2D grid. Furthermore, we demonstrate that the proposed approach can be used to compress static point cloud attributes efficiently.

\section{Related Work}
\label{sec:related}

Our work is at the crossroads of static point cloud attribute compression and deep representation learning of 3D data.
Compressing static point cloud attributes has been explored using graph transforms \cite{zhang_point_2014}, the Region-Adaptive Hierarchical Transform (RAHT) \cite{de_queiroz_compression_2016} and volumetric functions \cite{krivokuca_volumetric_2018}.
Graph transforms take advantage of the Graph Fourier Transform (GFT) and the neighborhood structure present in the 3D space to compress point cloud attributes.
The RAHT is a hierarchical transform which extends the Haar wavelet transform to an octree representation.
In this paper, we propose a different perspective, and leverage the manifold interpretation of the point cloud by mapping its attributes onto a 2D grid, which can then be compressed as an image. 

\begin{figure*}[t]
\centering
\begin{tikzpicture}[>=latex']
	\tikzset{fontsize/.style= {font=\footnotesize},
	block/.style= {fontsize,draw, fill=white, rectangle, align=center,minimum width=3cm,minimum height=0.9cm},
	mblock/.style={block, double copy shadow={shadow xshift=.5ex,shadow yshift=-.75ex,draw=black!30 }},
	node/.style= {fontsize,align=center,minimum width=2.5cm,minimum height=0.9cm}
	},
	\node [node]  (oriatt) {Original \\ attributes};
	\node [mblock, right=4.28cm of oriatt] (mapp) {Optimized Mapping};
	\node [mblock, right =.75cm of mapp] (imco) {Image \\ compression};
	\node [coordinate, below=0.9cm of imco] (immid) {};
	\node [node, above right=-0.46cm and 1.5cm of immid] (imbit) {Compressed \\ attributes};
	\node [mblock, below =1.8cm of imco] (imdc) {Image \\ decompression};
	\node [mblock, left =.75cm of imdc] (imap) {Inverse \\ mapping};
	\node [node, left =4.28cm of imap] (decatt) {Decompressed \\ attributes};
	\path[draw,<-] (imbit) edge (immid);

	\node [coordinate, below left =.901cm and 1.5cm of oriatt] (left) {};
	\node [coordinate, below right =.901cm and 2.7cm of imco] (right) {};
	\node [fontsize,above right = 0.75cm and 0cm of left] (enc) {\it Encoder};
	\node [fontsize,below right = 0.75cm and 0cm of left] (enc) {\it Decoder};
	\path[draw, dotted] (left) edge (right);

	\node [node, below=.44cm of oriatt]  (origeo) {Coded \\ geometry};
	\node [block, text=gray, draw=gray, right =.75cm of origeo] (segm) {Segmentation \\ into patches};
	\node [mblock, below=.44cm of mapp] (fold) {Grid Folding, \\ Folding refinement};

	\path[draw,->] (oriatt) edge (mapp)
		(mapp) edge (imco)
		(imdc) edge (imap)
		(imap) edge (decatt)

		(origeo) edge (segm)
		(segm) edge (fold)
		(fold) edge (mapp)
		(fold) edge (imap)
	;
	\path[draw,-] (imco) edge (immid);
	\path[draw,<-] (imdc) edge (immid);
\end{tikzpicture}
\caption{Proposed system for attribute compression. Segmentation is optional and can help to adapt to local geometry complexity.}
\label{fig:compsystem}
\end{figure*}
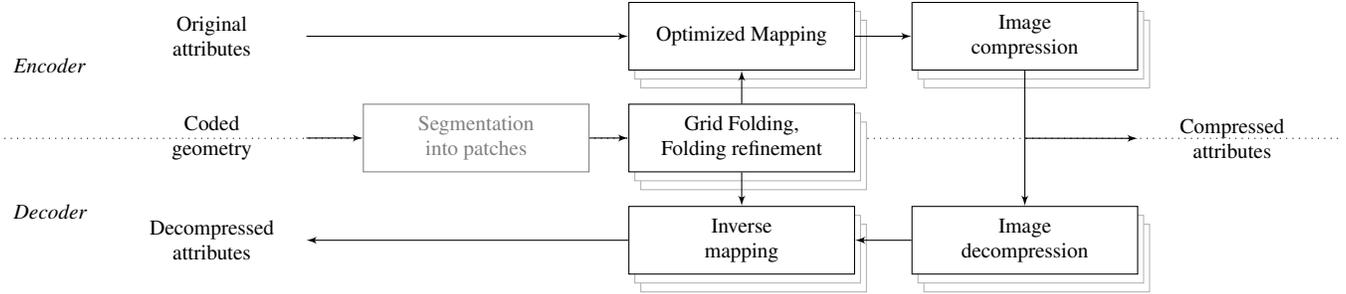

Deep learning methods have been used for representation learning and compression of point clouds \cite{quach_learning_2019}. 
In particular, the initial folding in our work is inspired by \cite{yang_foldingnet:_2017} where an autoencoder network is trained on a dataset to learn how to fold a 2D grid onto a 3D point cloud.
In our work, we build on this folding idea; however, we employ it in a very different way.
Specifically, we do not aim at learning a good representation that can generalize over a dataset; instead, we employ the folding network as a parametric function that maps an input 2D grid to points in 3D space.
The parameters of this function (i.e., the weights of the network) are obtained by overfitting the network to a specific point cloud.
In addition, the original folding proposed in~\cite{yang_foldingnet:_2017} is highly inefficient for PCC as it poorly adapts to complex geometries.
In our work, we propose a number of solutions to improve folding.

\section{Proposed method}
\label{sec:proposed}

\definetrim{philtrim}{8cm 0 8cm 0}
\begin{figure*}[htb]
\centering
\begin{subfigure}[t]{.24\textwidth}
    \vskip 0pt
    \centering
    \includegraphics[width=\linewidth, philtrim]{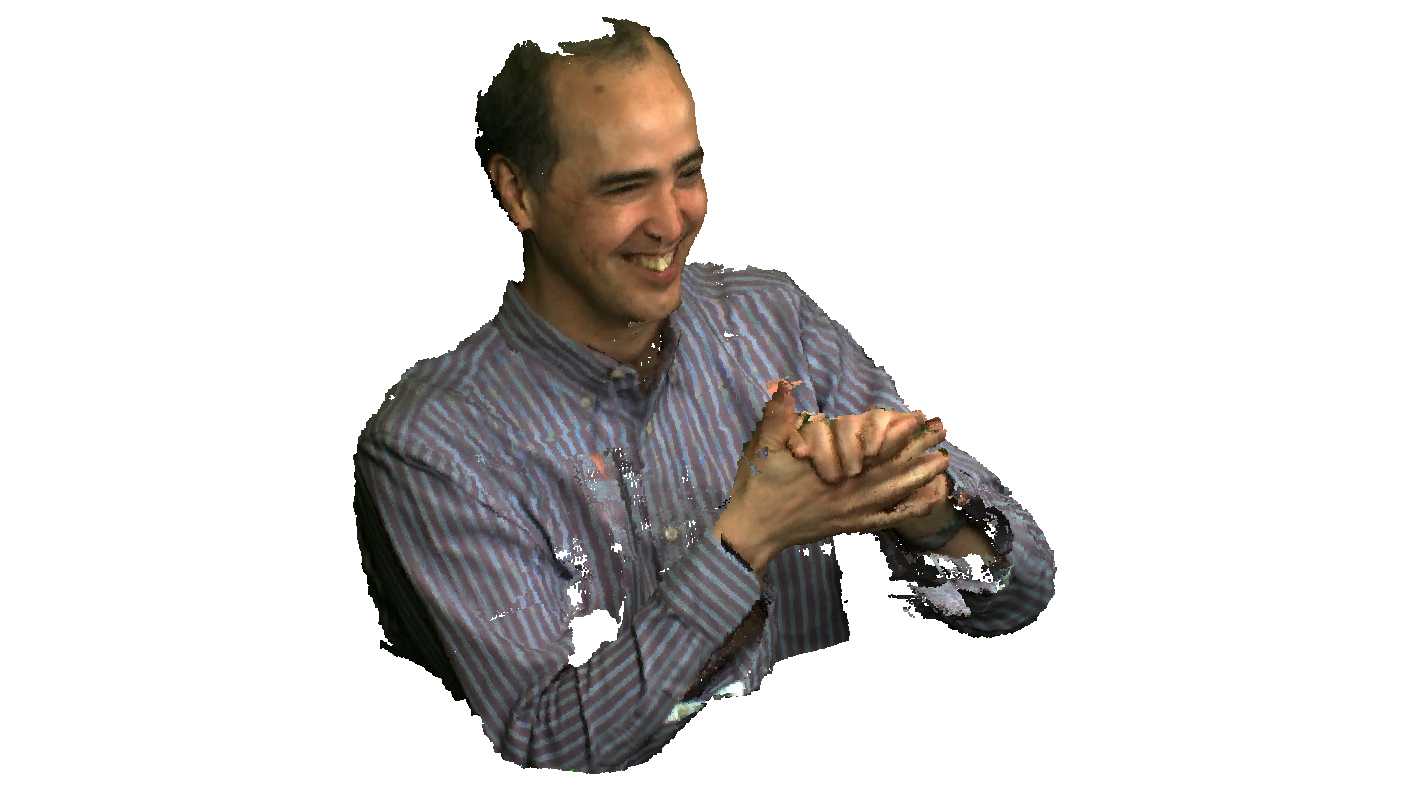}
\end{subfigure}
\begin{subfigure}[t]{.24\textwidth}
    \vskip 0pt
    \centering
    \includegraphics[width=\linewidth, philtrim]{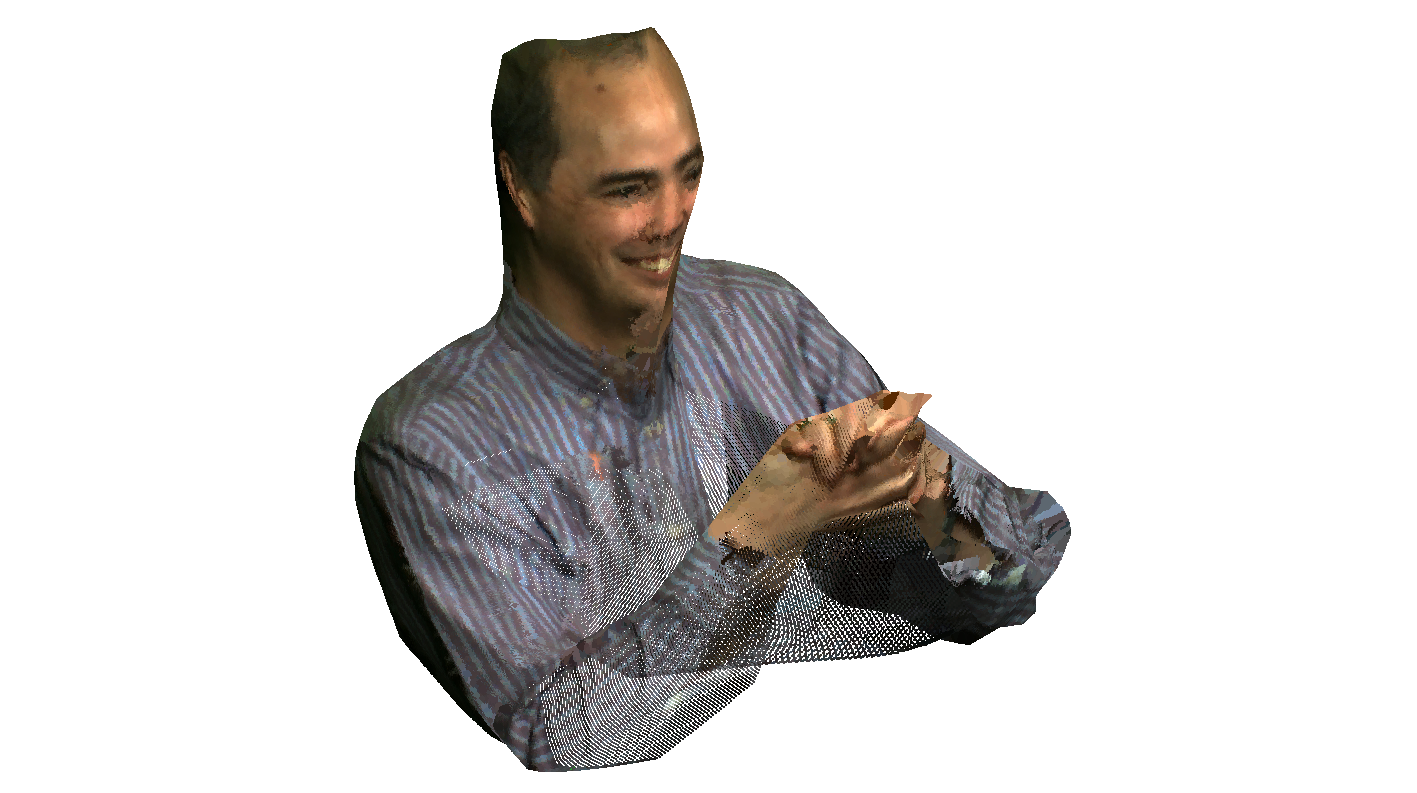}
\end{subfigure}
\begin{subfigure}[t]{.24\textwidth}
    \vskip 0pt
    \centering
    \includegraphics[width=\linewidth, philtrim]{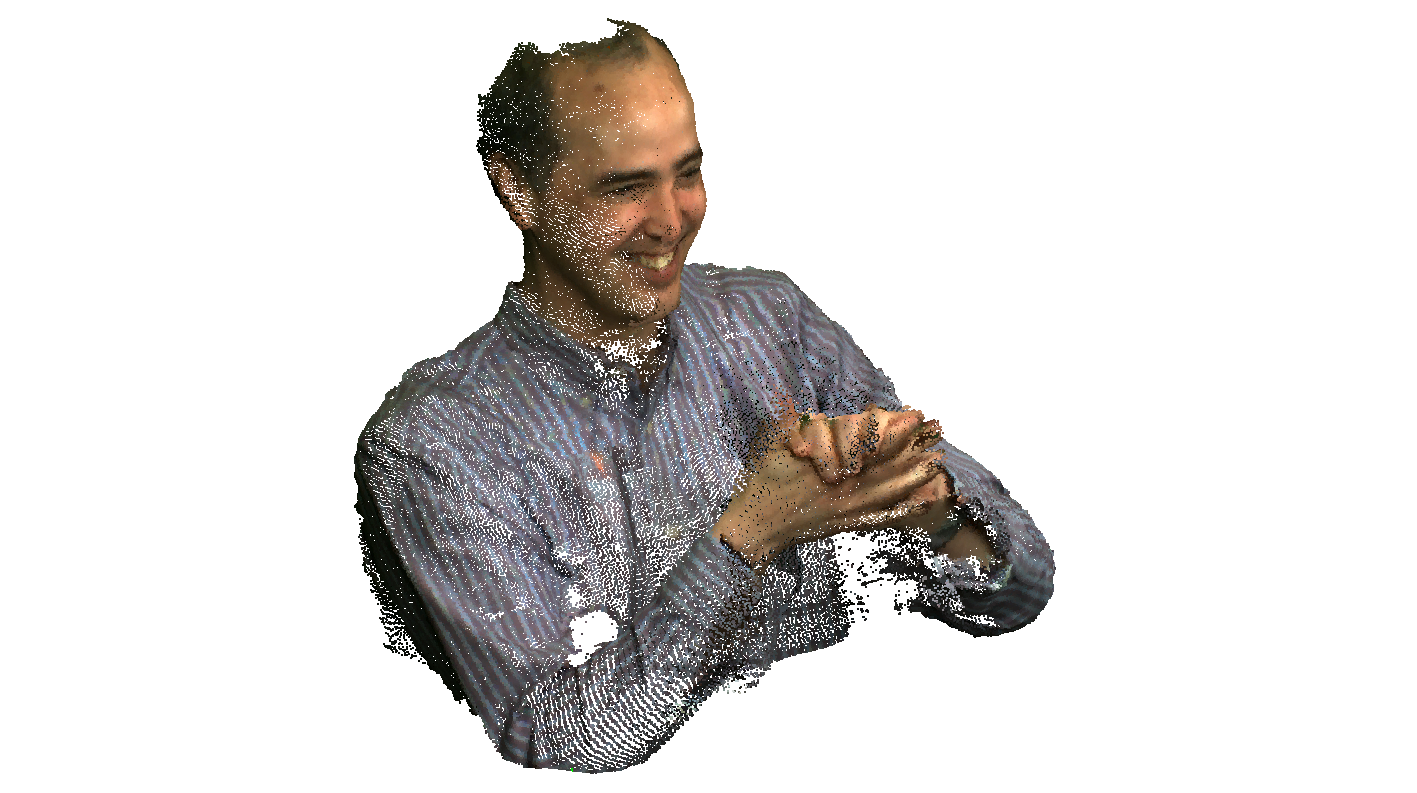}
\end{subfigure}
\begin{subfigure}[t]{.24\textwidth}
    \vskip 0pt
    \centering
    \includegraphics[width=\linewidth, philtrim]{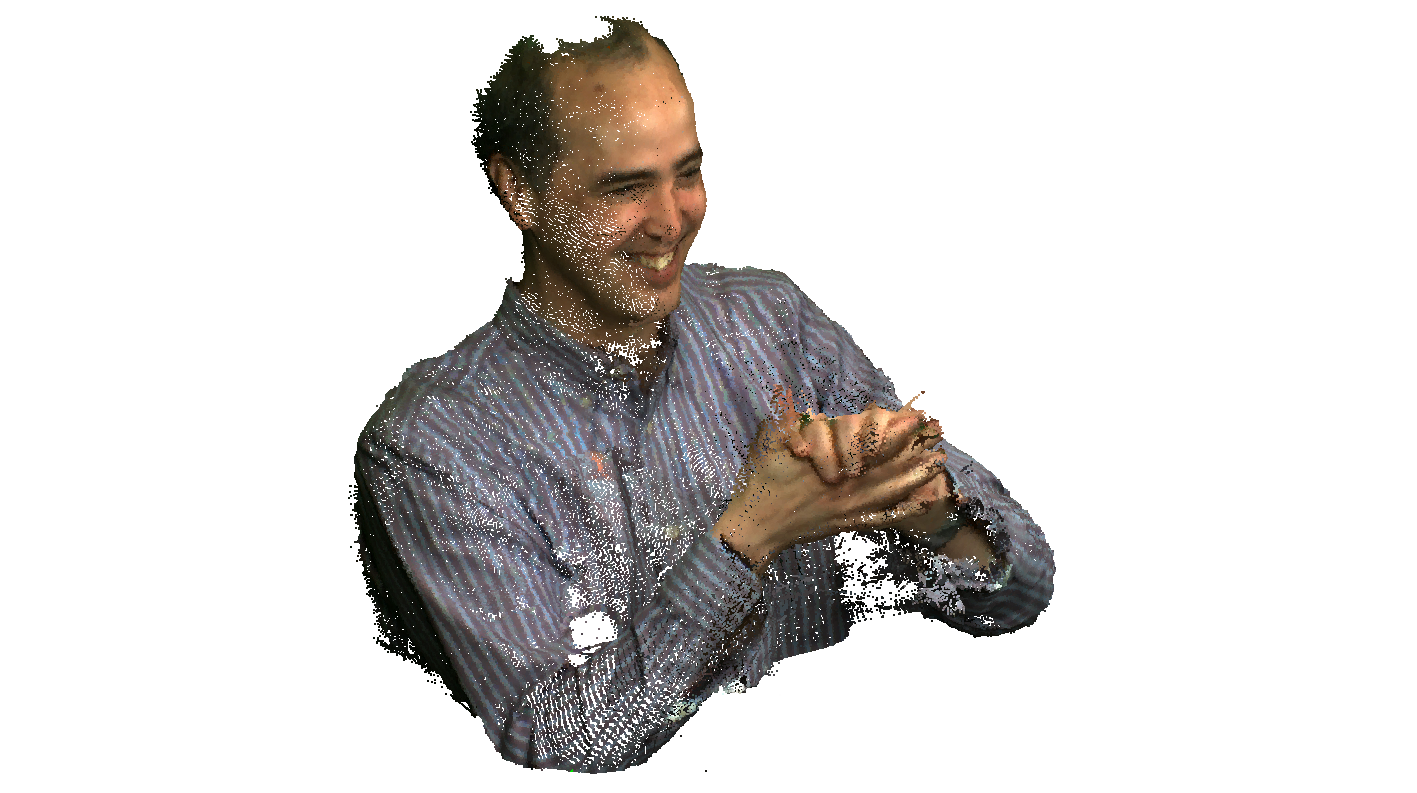}
\end{subfigure}
\begin{subfigure}[b]{.24\textwidth}
    \centering
    \includegraphics[width=0.73\linewidth]{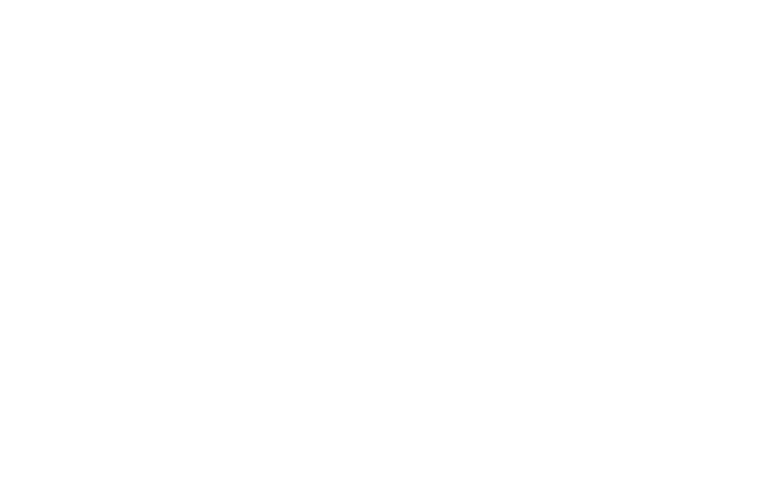}
    \caption{Original}
    \label{fig:foldingexampleoriginal}
\end{subfigure}
\begin{subfigure}[b]{.24\textwidth}
    \centering
    \includegraphics[width=0.73\linewidth]{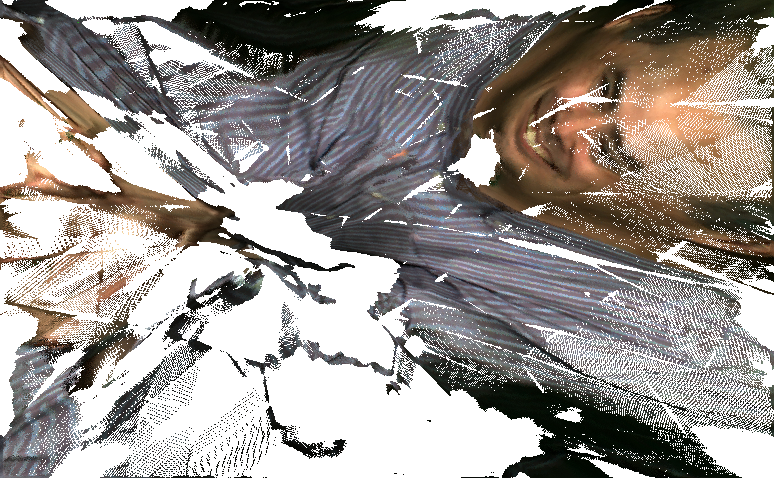}
    \caption{Folded ($27.63$ dB)}
    \label{fig:foldingexamplefolded}
\end{subfigure}
\begin{subfigure}[b]{.24\textwidth}
    \centering
    \includegraphics[width=0.73\linewidth]{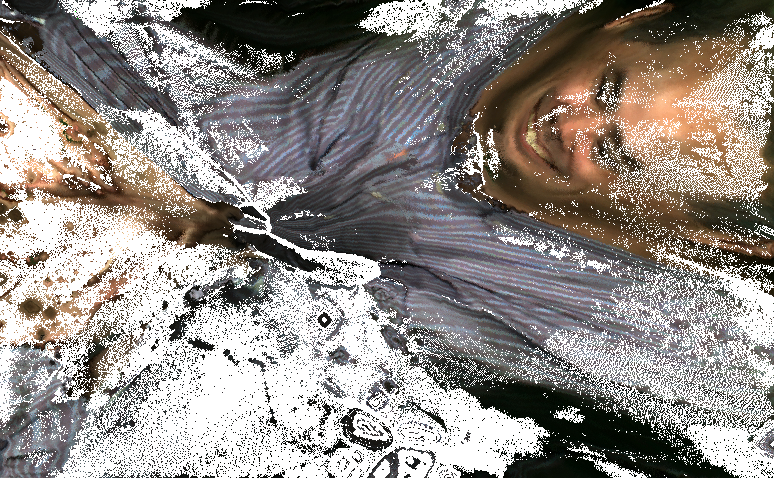}
    \caption{Refined folded ($30.62$ dB)}
    \label{fig:foldingexamplerefinedfolded}
\end{subfigure}
\begin{subfigure}[b]{.24\textwidth}
    \centering
    \includegraphics[width=\linewidth]{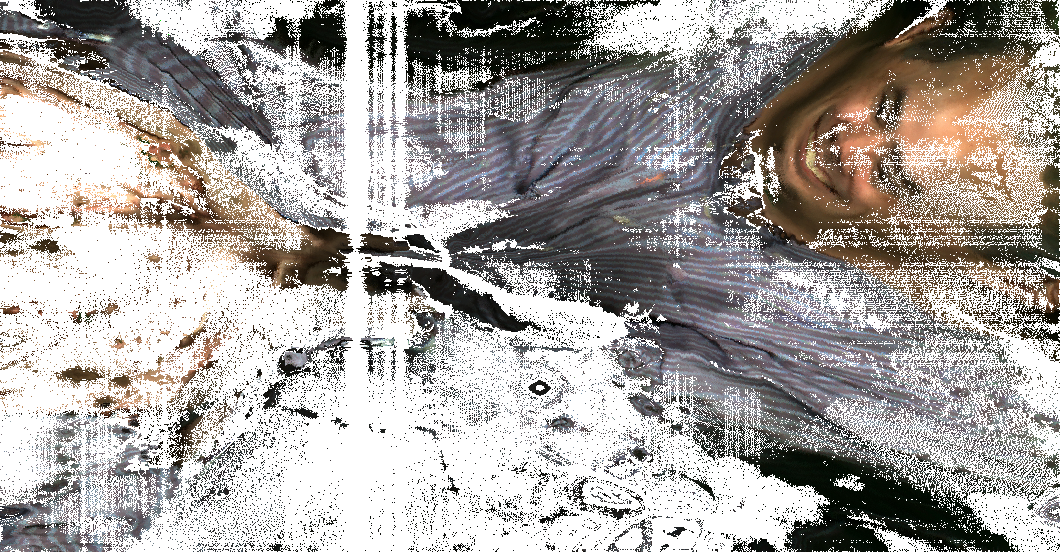}
    \caption{Opt. refined folded ($33.39$ dB)}
    \label{fig:foldingexampleoptimizedrefinedfolded}
\end{subfigure}
\caption{Different steps of our proposed attribute mapping method for  the first frame of phil9 \cite{loop_microsoft_2016}. Top row: phases of point cloud reconstruction; bottom row: the attributes mapped on a 2D grid, which is later compressed and transmitted. The initial folding (b) provides a rough reconstruction $\Xtilde$ which is improved with folding refinement (c) and occupancy optimization (d) to reduce the density mismatch between $\X$ and $\Xtilde$. We then map attributes from the point cloud onto a 2D grid. The holes in the grid are filled to facilitate compression with HEVC. We indicate Y PSNRs between original and colors distorted by mapping.}
\label{fig:foldingexample}
\end{figure*}

We propose a novel system for compressing point cloud attributes based on the idea that a point cloud can be seen as a discrete 2D manifold in 3D space.
In this way, we can obtain a 2D parameterization of the point cloud and we can map attributes from a point cloud onto a grid, making it possible to employ 2D image processing algorithms and compression tools.
The overall system is depicted in Figure~\ref{fig:compsystem}.
In a nutshell, our approach is based on the following two steps: a) we find a parametric function (specifically, a deep neural network) to fold a 2D grid onto a 3D point cloud; b) we map attributes (e.g., colors) of the original point cloud to this grid.
The grid and the parametric function contain all the necessary information to recover the point cloud attributes.
Assuming the point cloud geometry is coded separately and transmitted to the decoder, the folding function can be constructed at the decoder side, and the 2D grid is fully decodable without any need to transmit network parameters.
In practice, the 3D-to-2D mapping is lossy, which entails a mapping distortion in the step b) above. 
In the following, we propose several strategies to reduce this mapping distortion.

\textit{Notation.}
We use lowercase bold letters such as $\x$ to indicate 3D vectors (point cloud spatial coordinates), and uppercase letters such as $\X$ to indicate sets of 3D points (vectors). We denote with a tilde (like $\xtilde$ or $\Xtilde$) compressed (distorted) vectors or sets of vectors.
We use the notation $\avg{S} = \sum_{x \in S} x / \card{S}$ for the average over a set $S$.

\subsection{Grid folding}

We propose a grid folding composed of two steps, namely, an initial folding step to get a rough reconstruction of $\X$ and a folding refinement step to improve the reconstruction quality, which is quintessential to map point cloud attributes with minimal mapping distortion.

We fold a grid onto a point cloud to obtain its 2D parameterization by solving the following optimization problem:
\begin{equation}
	\min_{f} \Loss(\X, \Xtilde)
\end{equation}
where $\X$ is the set of $\n$ points in the original point cloud, $\Xtilde = \f(\X, \G)$ is the set of $\nprime$ points in the reconstructed point cloud obtained by folding $\G$ onto $\X$ where $\G$ the set of $\nprime = \w\times \h$ points of a 2D grid with 3D coordinates. In general, $\nprime \neq n$; however, we choose $\nprime$ to be close to $n$.
$\Loss$ is a loss function and $\f$ is a folding function.

We parameterize $\f$ using a neural network composed of an encoder $\fe$ and a decoder $\fd$ such that $\y = \fe(\X)$ and $\Xtilde = \fd(\G, \y)$.
The encoder $\fe$ is composed of four pointwise convolutions with filter sizes of $128$ followed by a maxpooling layer.
The decoder $\fd$ is composed of two folding layers with $\fd(\G, \y) = \FL(\FL(\G, \y), \y)$.
Each folding layer has two pointwise convolutions with filter sizes of $64$ and concatenates $\y$ to its input.
The last pointwise convolution has a filter size of $3$.
We use the ReLU activation~\cite{nair_rectified_2010} for the encoder and LeakyReLU activation~\cite{maas_rectifier_2013} for the decoder.
A one-to-one mapping exists between each point $\xtilde_i$ in the folded grid $\Xtilde$ and their original position $\g_i$ in the grid $\G$.

We propose the following loss function
\begin{equation}
	\Loss(\X, \Xtilde) = \dch(\X, \Xtilde) + \drep(\Xtilde)
\end{equation}
where $\dch$ is the Chamfer distance:
\begin{align*}
	\dch(\X, \Xtilde) &= \sum_{\x \in \X} \min_{\xtilde \in \Xtilde} \norm{\x - \xtilde}^{2} + \sum_{\xtilde \in \Xtilde} \min_{\x \in \Xtilde} \norm{\xtilde - \x}^{2},
\end{align*}
and $\drep$ is a novel \textit{repulsion loss} computed as the variance of the distance of each point in $\Xtilde$ to its nearest neighbor:
\begin{align*}
	\drep(\Xtilde) &= \Var(\set{\min_{\xtilde^\prime \in \Xtilde \setminus \xtilde} \norm{\xtilde - \xtilde^\prime}^{2} | \xtilde \in \Xtilde}).
\end{align*}
The Chamfer distance ensures that the reconstruction $\Xtilde$ is similar to $\X$ and the repulsion loss penalizes variations in the reconstruction's density.

We obtain the parameterized folding function $\f$ by training a neural network using the Adam optimizer \cite{kingma_adam:_2014}.
We use the point cloud $\X$ as the single input which is equivalent to overfitting the network on a single sample.

\subsection{Folding refinement}

The initial folding has difficulties reconstructing complex shapes accurately as seen in Figure \ref{fig:foldingexamplefolded}.
Specifically, the two main issues are mismatches in local density between $\X$ and $\Xtilde$ and inaccurate reconstructions for complex shapes.
As a result, this introduces significant mapping distortion when mapping attributes from the original PC to the folded one; additionally, this mapping distortion affects the reconstructed point cloud attributes.
For compression applications, this is a serious issue as there are now two sources of distortion from both mapping and compression.
This is why we propose a folding refinement method that alleviates mismatches in local density and inaccurate reconstructions.

First, we reduce local density variations by considering density-aware grid structure preservation forces inside $\Xtilde$.
Specifically, each point $\xtilde$ is attracted towards the inverse density weighted average of its neighbors $\pgrid$.
Since a one-to-one mapping exists between $\Xtilde$ and $\G$, each point $\xtilde_i$ in the folded grid $\Xtilde$ has a corresponding point $\g_i$ in the grid $\G$.
We then define the inverse density weight $\idw_i$ for $\xtilde_i$ as $\idw_i = \avg{\set{\norm{\xtilde_i - \xtilde_j} | \g_j \in \N{\G}{\g_i}}}$ with $\N{\G}{\g_i}$ the set of horizontal and vertical neighbors of $\g_i$ in the grid $\G$.
This encourages the reconstruction to have a more uniform distribution by penalizing high density areas.
Given the set $\Idw$ comprising all weights $\idw_i$, we define the normalized weights $\idwn_i = (\idw_i - \min(\Idw))/(\max(\Idw) - \min(\Idw))$.
Finally, this allows us to define the weighted average $\pgrid[_i] = \avg{\set{\idwn_{j}\xtilde_{j} | \g_{j} \in \N{\G}{\g_{i}}}}$.

\begin{figure*}[t]
\centering
\begin{subfigure}{.33\textwidth}
\centering
\includegraphics[width=\linewidth]{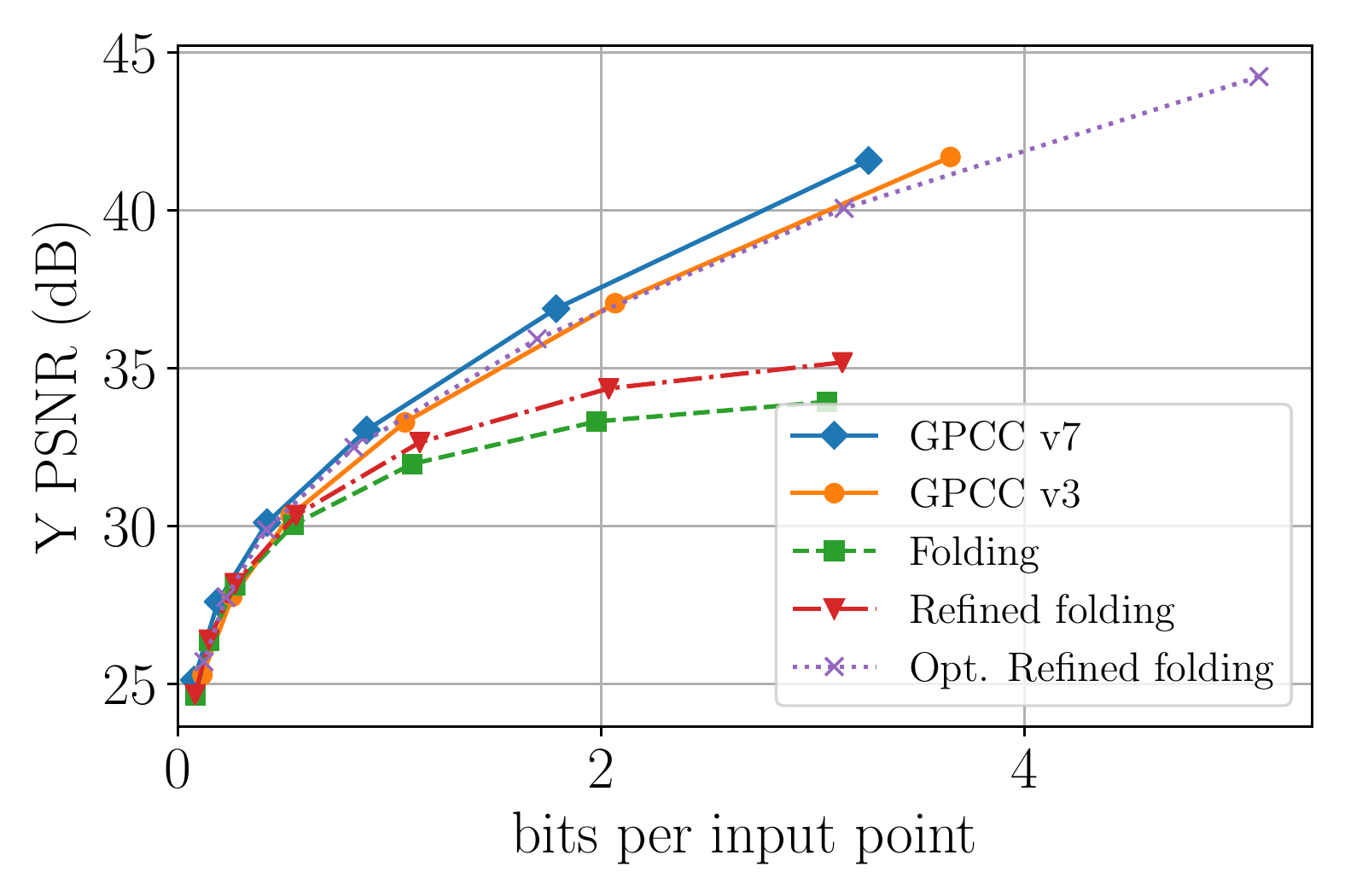}
\end{subfigure}
\begin{subfigure}{.33\textwidth}
\centering
\includegraphics[width=\linewidth]{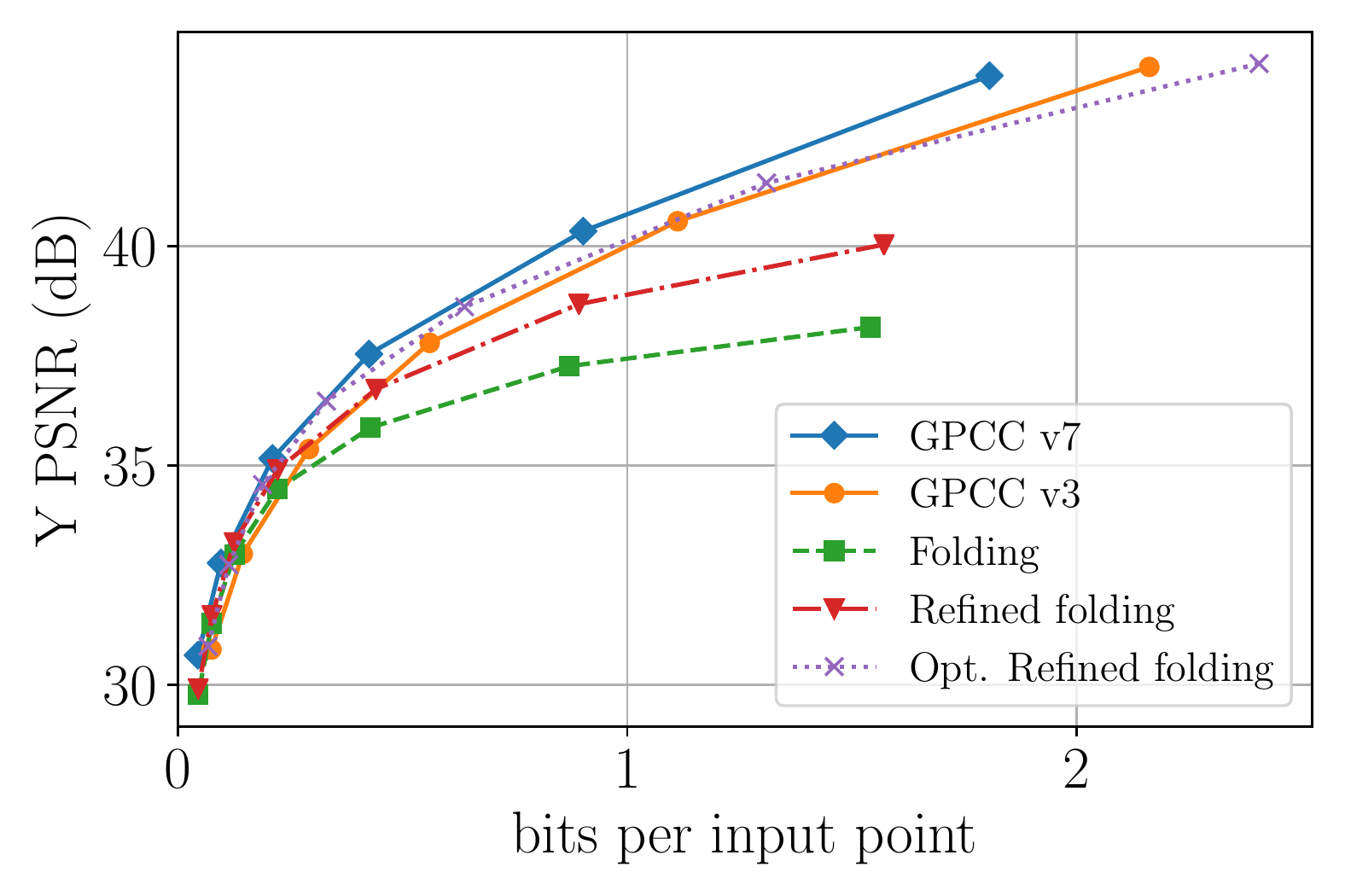}
\end{subfigure}
\begin{subfigure}{.33\textwidth}
\centering
\includegraphics[width=\linewidth]{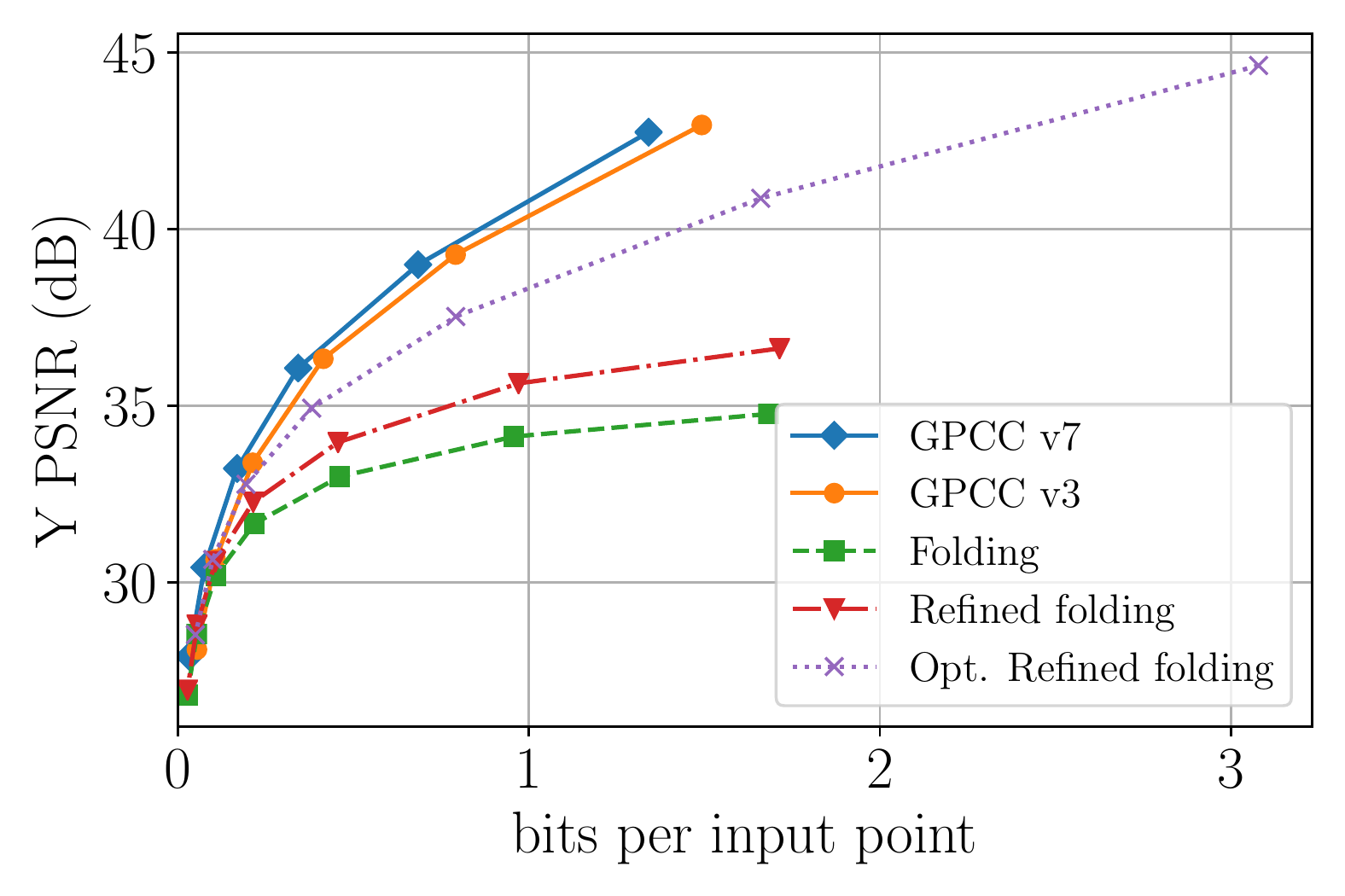}
\end{subfigure}
\caption{RD curves showing the performance of the different steps of our method. From top to bottom: longdress\_vox10\_1300, redandblack\_vox10\_1550 and soldier\_vox10\_0690 \cite{schwarz_common_2018}.}
\label{fig:rdcurves}
\end{figure*}

Second, we set up bidirectional attraction forces between $\X$ and $\Xtilde$ to solve two issues: incomplete coverage, when $\Xtilde$ does not cover parts of $\X$, and inaccurate reconstructions, when $\Xtilde$ fails to reproduce $\X$ accurately.
As a solution, we attract each point $\xtilde$ towards two points $\ppush$ and $\ppull$.
Specifically, $\ppush$ is the nearest neighbor of $\xtilde$ in $\X$ and \textit{pushes} $\Xtilde$ towards $\X$ which allows for more accurate reconstructions.
On the other hand, $\ppull$ is the average of the points in $\X$ which have $\xtilde$ as their nearest neighbor and allows $\X$ to \textit{pull} $\Xtilde$ closer which alleviates incomplete coverage issues.

Finally, we combine these components into an iterative refinement system to update the point cloud reconstruction:
\begin{equation}
  \xtilde_{t+1,i} = \alpha \pgrid[_{t,i}] + (1-\alpha)(\ppush[_{t,i}] + \ppull[_{t,i}]) / 2
\end{equation}
where $\xtilde_{t,i}$ is the value of $\xtilde_i$ after $t$ iterations and $\xtilde_{0} = \xtilde$.
The inertia factor $\alpha \in [0, 1]$ balances the grid structure preservation forces in $\Xtilde$ with the bidirectional attraction forces set up between $\X$ and $\Xtilde$.
Preserving the grid structure preserves the spatial correlation of the attributes mapped on the grid and the density-aware aspect of these forces results in more uniformly distributed points.
In addition, the bidirectional forces improve the accuracy of the reconstruction significantly.

\subsection{Optimized Attribute Mapping}
Once a sufficiently accurate 3D point cloud geometry is reconstructed (Figure \ref{fig:foldingexamplerefinedfolded}), we can map attributes from $\X$ to $\Xtilde$.
To this end, we first build a mapping $\mXXt$ from each point in $\X$ to a corresponding point in $\Xtilde$ (for example, the nearest neighbor). 
Hence, the inverse mapping $\mXtX$ maps $\xtilde$ back to $\X$. As $\mXXt$ is not one-to-one (due to local density mismatches and inaccuracy of the reconstruction), several points in $\X$ can map to the same $\xtilde$. Thus, a given $\xtilde$ can correspond to zero, one or many points in $\X$; we define the number of these points as its occupancy $\occ(\xtilde)$. 
Attribute mapping from $\X$ to $\Xtilde$ is obtained using $\mXtX$ as the attribute value for a point $\xtilde$ is the average of the attribute values of $\mXtX(\xtilde)$.
In case $\mXtX(\xtilde)=\varnothing$, we simply assign to $\xtilde$ the attribute of its nearest neighbor in $\X$.
As a consequence of this approach, points with higher occupancy tend to have higher mapping distortion, as more attributes are averaged.

To overcome this problem, we integrate the occupancy as a regularizing factor when building the mapping.
For each point $\x$ in $\X$, we consider its $k$ nearest neighbors set $\ngb_{k}(\x) \in \Xtilde$ and select $\mXXt(\x) = \argmin_{\xtilde \in \ngb_{k}(\x)} \occ(\xtilde)\norm{\xtilde - \x}$.
Specifically, the mapping is built iteratively and the occupancies are updated progressively.

As noted above, when $\occ(\xtilde) > 1$, the attributes are averaged which introduces distortion.
We mitigate this problem by adding rows and columns in the 2D grid (see Fig.~\ref{fig:foldingexampleoptimizedrefinedfolded}) using the following procedure.
Since $\occ(\xtilde)$ is defined on $\Xtilde$ and there is a one-to-one mapping between $\Xtilde$ and $G$, we can compute mean occupancies row-wise and column-wise.
In particular, we compute mean occupancies with zeros excluded and we select the row/column with the maximum mean occupancy.
Then, we reduce its occupancy by inserting additional rows/columns around it.
We repeat this procedure until we obtain a lossless mapping or the relative change on the average of mean occupancies $\Delta_{r}$ is superior to a threshold $\Delta_{r,min}$.

\section{Experimental results}
\label{sec:experimental}

We evaluate our system for static point cloud attribute compression and compare it against G-PCC v3 \cite{mammou_pcc_2018} and v7 \cite{noauthor_g-pcc_2019}.
We also study the impact of folding refinement and occupancy optimization on our method by presenting an ablation study.
Since folding is less accurate on complex point clouds, we manually segment the point clouds into patches and apply our scheme on each patch. The patches are then reassembled in order to compute rate-distortion measures.

We use TensorFlow 1.15.0 \cite{abadi_tensorflow_2016}.
For the folding refinement, we set $\alpha$ to $1/3$ and perform 100 iterations.
When mapping attributes, we consider $k=9$ neighbors for assignment.
When optimizing occupancy, we set $\Delta_{r,min}$ to $10^{-6}$.
We then perform image compression using BPG \cite{bellard_bpg_nodate}, an image format based on HEVC intra \cite{noauthor_high_nodate}, with QPs ranging from $20$ to $50$ with a step of $5$.

In Figure \ref{fig:rdcurves}, we observe that our method performs comparably to G-PCC for ``longdress" and ``redandblack".
The performance is slightly worse for ``soldier" as its geometry is much more complex making a good reconstruction difficult and introducing mapping distortion.
We obtain significant gains in terms of rate-distortion by improving the reconstruction quality using folding refinement and occupancy optimization.
This shows the potential of our method and confirms the importance of reducing the mapping distortion.

\section{Conclusion}
\label{sec:conclusion}
Based on the interpretation of a point cloud as a 2D manifold living in a 3D space, we propose to fold a 2D grid onto it and map point cloud attributes into this grid.
As the mapping introduces distortion, this calls for strategies to minimize this distortion.
In order to minimize mapping distortion, we proposed a folding refinement procedure, an adaptive attribute mapping method and an occupancy optimization scheme.
With the resulting image, we compress point cloud attributes leveraging conventional image codecs and obtain encouraging results.
Our proposed method enables the use of 2D image processing techniques and tools on point cloud attributes.

\bibliographystyle{IEEEbib}
\bibliography{main}

\end{document}